\documentclass [12pt]{article}
\usepackage{amsmath,amsfonts}

\begin{document}
\begin{center}
{\large\bf The critical manifold of the Lorentz--Dirac equation\vspace{2cm}}\\
{HERBERT SPOHN} \medskip\\
{\it Zentrum Mathematik and Physik Department, \\
TU M\"unchen, D--80290 M\"{u}nchen, Germany}
 {\tt spohn@mathematik.tu-muenchen.de}
\end{center}\bigskip 
PACS.03.50.De\\
PACS.41.20.-q\vspace{1.5cm}\\
{\bf Abstract}. We investigate the solutions to the
Lorentz--Dirac equation and show that its solution flow has a structure
identical to the one of
renormalization group flows in critical phenomena. The physical
solutions of the Lorentz--Dirac equation lie on the critical
surface. The critical surface is repelling, i.e. any slight
deviation from it is amplified and as a result the solution runs away to
infinity. On the other hand, Dirac's asymptotic condition (acceleration 
vanishes for
long times) forces the solution to be on the critical manifold.
The critical surface can be determined perturbatively. Thereby one
obtains an effective second order equation, which we apply to various
cases, in particular to the motion of an electron in a Penning 
trap.\vspace{1.5cm}\\

The Lorentz--Dirac equation governs the motion of a classical
charge in prescribed external electromagnetic fields and including
radiation reaction, i.e. the loss of energy due to radiation. In
standard relativistic notation it reads \cite{R}
\begin{equation}\label{1}
m \dot{v}^\mu = e F^{\mu\nu}(x)v_\nu +(e^2/ 6
\pi c^3) [\ddot{v}^\mu - \frac{1}{c^2}\,\dot{v}^\lambda
\dot{v}_\lambda v^\mu ] \, .
\end{equation}
Here $m$
is the (experimental) rest mass of the particle with charge $e$.
$x^{\mu}(s)$ is the world line and
$v^{\mu}(s)= \dot{x}^{\mu}(s)$
the velocity of the charge parametrized in
its eigentime $s$.  The
particle is subject to time--independent external fields as given through
the electromagnetic field tensor $F^{\mu\nu}$. The first term in
(1) is the Lorentz force while the second term describes the
radiation reaction.

One obvious issue is to understand how the Lorentz--Dirac equation
is related to the Maxwell--Lorentz equations with a suitable
ultraviolet cut--off. This problem was studied extensively by
Abraham, Lorentz, and many others, cf. \cite{R,Y} for a detailed
account. In his famous paper \cite{D}, Dirac circumvented the
issue through a somewhat delicate splitting of the
fields generated by a point charge.  The, to our knowledge most
complete formal derivation of (\ref{1}) has been worked out by Nodvik
\cite{N}. Some rigorous results are \cite{KKS,KS,KS1}. For the purpose
of this letter we regard the Lorentz--Dirac equation as given.

As noted already by Dirac, Eq. (\ref{1}) has runaway solutions which
grow exponentially in time, simply because for $F^{\mu\nu} = 0$ and in
the approximation of small velocities we have
 $m \dot{{\mathbf  v}} =
(e^2 / 6\pi c^3) \ddot{{\mathbf  v}}$. Dirac \cite{D}, reemphasized by Haag
\cite{H}, postulated that the physical solutions to (\ref{1}) must
satisfy the asymptotic condition
$\lim\limits_{s\to \infty} \, \dot{v}^{\mu} (s)=0,$
which, as extra bonus, is a substitute for the missing initial condition
 $\ddot{{\mathbf x}}(0)$. The validity of the asymptotic condition has been
tested only in explicit cases \cite{R,B,Ga}. With a general external
field tensor $F^{\mu\nu}$ the solution behavior of (\ref{1}) might be
complicated and should expected to be chaotic. Physical and
unphysical solutions might be thoroughly mixed. Thus in
principle, for given ${\mathbf  x}(0), \dot{{\mathbf  x}}(0)$,
there could be  many
solutions satisfying the asymptotic condition. Which one to pick
then? On a more practical level, one would like to have a reliable 
numerical scheme not hampered by the instability of physical solutions.

The purpose of this letter is to explain that the solution flow of
the Lorentz--Dirac equation has a structure familiar from the
renormalization group flows in critical phenomena. The physical
solutions lie on the critical surface, which contains attractive
fixed points whose location depends on $F^{\mu\nu}$. Slightly off the
critical surface, the solution grows exponentially fast, so to
speak it flows to the high, resp. low, temperature fixed point.
Our observation has two important implications. (1) The critical
manifold is actually a surface of the form
$\ddot{{\mathbf  x}} = h({\mathbf  x},
\dot{{\mathbf  x}})$. Thus, for given initial conditions
${\mathbf  x}(0), \dot{{\mathbf  x}}(0)$,
there is exactly one solution on the critical surface and, as to
be shown, it satisfies the asymptotic condition. (2) There is an
effective second order equation, given below, which governs the
motion on the critical surface. Thus the initial value problem is
restored and the equation can esasily be
solved numerically. We will demonstrate the predictive power of the
second order equation by a few examples, still handled
without numerical integration, the physically most relevant
of which is the motion of an
electron in a Penning trap \cite{BG}. (This system was pointed out to
us by Wolfgang Schleich).

In all applications the radiation reaction is a small correction
to the Lorentz force equation, which means that
the radiation reaction term, the {\it highest}
derivative in (\ref{1}), carries a small prefactor.
Differential equations of such a type have been studied
extensively through singular (or geometric) perturbation theory
\cite{KSa,J}, which is closely connected to the theory of center
manifolds. The application to the Lorentz-Dirac equation is a little
bit messy and has been carried out in \cite{KS1}.
Rather than trying to summarize these results, we believe it to be 
more instructive to
illustrate the basic features of the method by using a
fictituous mathematical example.

Let us consider then the ordinary differential equation of the form
\begin{equation}\label{2}
\dot{x} = g(x,y)\, , \quad \varepsilon \dot{y} =y-h(x)\, ,
\end{equation}
$x(t)\in {\mathbb R}$, $y(t) \in {\mathbb R}$. We want to understand
the behavior of solutions  for small $\varepsilon$.
If we simply set $\varepsilon = 0$, the second equation reduces to $y=h(x)$
and therefore $\dot{x} = g(x, h(x))$. The ambient phase space has
disappeared and the motion takes place only on the one--dimensional
surface $\{ y=h(x)\}$. On the other hand we can go over to the
slow time scale $\tau, \tau=\varepsilon^{-1} t$. Denoting
differentiation with respect to $\tau$ by $^\prime$, (\ref{2}) reads
\begin{equation}\label{3}
x^\prime = \varepsilon g(x,y)\, ,\quad y^\prime =y-h(x)\, .
\end{equation}
Setting now $\varepsilon = 0$, yields $x^\prime =0$, i.e.
$x(\tau)= x_0$ and $y^\prime=y-h(x_0)$. Thus the surface $\{ y=h(x)\}$
consists exclusively of repelling fixed points. If $y(0) \not= h(x_0)$,
the solution 
grows exponentially. In this sense the surface $\{ y=h(x)\}$ is critical.
The main result of geometric singular perturbation theory is
that for small $\varepsilon$ the critical surface persists and is
of the form $\{ y=h_\varepsilon(x)\}$. On the critical surface the
motion is governed by
\begin{equation}\label{3a}
\dot{x} = g(x,h_\varepsilon(x))\, .
\end{equation}
If $x(0),y(0)$ are off the critical surface, the
solution to (\ref{2}) diverges exponentially with rate $1/\varepsilon$.

Of course, abstractly only the existence of $h_{\varepsilon}$
is asserted. Its concrete form must be extracted from (\ref{2}).
Fortunately we are allowed to determine $h_{\varepsilon}$
perturbatively (which is not the case for individual solutions).
We make the
ansatz $ y=h_\varepsilon(x)= h_0(x) + \varepsilon h_1(x) +
{\cal O} (\varepsilon^2)$ and insert
 in (\ref{2}). This results in $\varepsilon \dot{y}=  \varepsilon  h_0^\prime(x)
 \dot{x} + {\cal O} (\varepsilon^2) = \varepsilon  h_0^\prime(x)
  g(x,h_0(x)) + {\cal O} (\varepsilon^2) =
 h_0(x)+\varepsilon h_1(x) - h(x)  + {\cal O} (\varepsilon^2)$.
 Comparing orders of $\varepsilon$ yields
 $ h_0(x) = h(x)$, $h_1(x) = h_{0}^\prime(x)g(x,h_{0}(x))$.
 Inserting in
 (\ref{3a}), we have the
 effective equation of motion
\begin{equation}\label{4}
\dot{x} = g(x,h(x)) + \varepsilon  \partial_y g(x, h(x))h^\prime(x)
g(x,h(x))\, ,
\end{equation}
valid up to an error of order $\varepsilon^{2}$.
 
Returning to the Lorentz--Dirac equation, by the same argument it
has a repelling critical surface of the form
$\{\ddot{{\mathbf  x}}=h({\mathbf  x},\dot{{\mathbf  x}})\}$.
 To
understand the motion on the critical surface we fix one inertial frame and
denote the position and velocity three--vectors by
${\mathbf  r}(t), {\mathbf  u}(t)$.
If $\phi$ is the external electrostatic potential, then the
energy balance reads
\begin{eqnarray}\label{5}
\frac{d}{dt} [\gamma mc^2 + e \phi({\mathbf  r}) - (e^2/6  \pi c^3)
\gamma^4 ({\mathbf  u}\cdot \dot{{\mathbf  u}})]\nonumber\\
=-(e^2/6 \pi c^3) [\gamma^4 \dot{{\mathbf  u}}^2 +
c^{-2}\gamma^6  ({\mathbf  u}
\cdot \dot{{\mathbf  u}})^2]
\end{eqnarray}
with $\gamma= 1 /\sqrt{1-{\mathbf  u}^2/c^2}$. We integrate both sides of
(\ref{5}) in time. Since on the critical
manifold the
Schott term $-(e^2/6 \pi c^3) \gamma^4
({\mathbf  u} \cdot \dot{{\mathbf  u}})$ is
bounded, we conclude
that
\begin{equation}\label{6}
\int _{0}^\infty dt [\gamma^4 \dot{{\mathbf  u}}^2 + c^{-2}\gamma^6
({\mathbf  u} \cdot \dot{{\mathbf  u}})^2] < \infty
\end{equation}
on the critical surface. This is possible only if the asymptotic condition
$\lim\limits_{t\to\infty} \dot{{\mathbf  u}}(t)=0$ holds. Off the critical 
manifold
$ \dot{{\mathbf  u}}(t)$
diverges. Thus given ${\mathbf  r}(0), \dot{{\mathbf  r}}(0)$, the asymptotic
condition
singles out the {\it unique} $\ddot{{\mathbf  r}}(0)$ on the critical surface.

Inserting the asymptotic condition in (\ref{1}),
we see that $-\nabla \phi({\mathbf  r}(t))
\to 0$ as $t \to \infty$, which implies in essence two distinct
scenarios. (i) The particle is scattered into a region where $F^{\nu\mu}=0.$
Then ${\mathbf  u}(t)$ has a limit as  $t \to \infty$ and ${\mathbf  r}(t)$ grows
linearly. (ii) The motion is bounded. Then the particle comes to
rest, $\lim\limits_{t\to\infty} {\mathbf  u}(t)=0$, at a point where the
electrostatic force, $-\nabla\phi$, vanishes. Note that in
general the condition $-\nabla\phi({\mathbf  r}_\infty)=0$ does not
determine the asymptotic position ${\mathbf  r}_\infty$. E.g. a uniform
magnetic field is confining even for $\phi({\mathbf  r})=0$.

In analogy to (\ref{4}), our next task is to derive an effective second
order equation for
the motion on the critical surface. We follow the steps
leading to (\ref{4}) and obtain
\begin{eqnarray}\label{7}
&&m\dot{v}^\mu = e F^{\mu\nu}({\mathbf  x})v_\nu + (e^2/6 \pi c^3) \{(e/m)
v^\sigma(\partial_\sigma F^{\mu\nu}({\mathbf  x}))v_\nu\nonumber\\
&&+ (e/m)^2 F^{\mu\alpha} ({\mathbf  x}) F_\alpha^{\;\;\nu}
({\mathbf  x})v_\nu + (e/mc)^2
F^{\sigma\alpha}({\mathbf  x})
F_\alpha^{\;\;\nu}({\mathbf  x}) v_\sigma v_\nu  v^\mu\}\, .
\end{eqnarray}
In principle one could compute also higher order terms. But they
have the same magnitude as those contributions neglected already
in the derivation of the Lorentz--Dirac equation. In addition
(\ref{7})  correctly describes the  long time behavior
as dominated by radiation reaction. Higher orders yield no
qualitative change and, at best, make a minute correction of relative 
order $10^{-24}$ or even smaller in concrete examples.

Eq. (\ref{7}) appears in the second volume of the course in
theoretical physics by  Landau and Lifshitz \cite{LL},
who were guided by the insight that radiation reaction must have
a small effect. One can only speculate why the Landau and
Lifshitz equation (\ref{7}) is apparently ignored in the literature.
For sure, they
do not discuss the structure of the flow with its critical
manifold nor the relation to the asymptotic condition.

There are several cases of interest where the solution to (\ref{7})
can still be handled analytically. The first one is a vanishing
magnetic field and an electrostatic potential varying only along
the 2--axis. Setting
${\mathbf  r}=(0,y,0)$,
${\mathbf  u}=(0,\dot{y},0)$,  $e\phi({\mathbf  r}) = V(y)$,
the one--dimensional motion is governed by
\begin{equation}\label{8}
\frac{d}{dt} \, (m\gamma \dot{y}) =- V^\prime(y) - (e^2/6\pi c^3)
(1/m) V^{\prime\prime}(y)\gamma\dot{y}\, .
\end{equation}
If $V$ is convex, the energy is damped monotonically. The particular 
case of a quadratic potential is studied in the recent third edition of 
the textbook 
by Jackson \cite{Ja} in the context of line breadth and level shift of a
radiating oscillator. However, if $V$
is
periodic, say $V(y) = V_{0}\cos(k_{0} y)$, then at the maxima the particle
gains in energy from the near field, a process dominated
by the energy loss at the minima. For long times the particle comes to
rest. Also if $V$ has a linear piece, then in this spatial interval 
the charge is accelerated without friction.

An experimentally more accessible set--up is the motion in a
uniform magnetic field $(0,0,B)$. Then (\ref{7}) simplifies to
\begin{equation}\label{8a}
\frac{d}{dt} \, (m\gamma {\mathbf  u}) = e B {\mathbf  u}^\bot- (e^2/6\pi c^3)
(e B/m)^2 \gamma^2 {\mathbf  u}
\end{equation}
with ${\mathbf  u}=(u_1,u_2,0) =
u(\cos \varphi,\sin \varphi,0)$,
${\mathbf  u}^\bot=(-u_2,u_1,0)$.
Setting $\alpha= e^2/6 \pi m c^3$ and the cyclotron frequency
$\omega_{c}=eB/m$, one obtains $du/d\varphi = - \alpha \omega_{c}
\varphi$, i.e. $u(\varphi)=u(0) \exp [-\alpha \omega_{c}\varphi]$.
For electrons $\alpha\omega_{c} = 8.8 \times 10^{-18} B[\mbox{Gauss}]$, which
shows that radiative reaction is very small even at high fields.
The charge spirals to its central rest point. Using (\ref{8a}) the
radius shrinks as
\begin{equation}\label{9}
r(t)=r_0 e^{-\alpha\omega_{c}^2 t}/(1+ (\gamma -1)((1- e^{-2 \alpha\omega_{c}^2
t})/2))\, .
\end{equation}
In the ultrarelativistic regime,  $\gamma \gg 1$, (\ref{9})
simplifies to $r(t) = r_0 (1+ \gamma \alpha\omega_{c}^2 t)^{-1}$,
provided $\alpha\omega_{c}^2t \ll 1$. To have some order of magnitude,
in the case of an electron, $\alpha\omega_{c}^2=1.6 \times10^{-6}
(B[\mbox{Gauss}])^2/\sec$
and $r_0=  1.7 \times 10^{-3}(\gamma/(B[\mbox{Gauss}]))$\,m.
Thus for $B=10^{3}$\,Gauss and an ultrarelativistic $\gamma=6 \times 
10^4$ the radius
shrinks within $0.9 \sec$  from
its initial value $r_0=10$\,cm to $r=1\,{\mu}$m, at which time the 
electron has made $2 \times 10^{14}$ revolutions. Only then
the power law, $(1+t)^{-1}$, crosses over to an exponential
damping.

Our third example is the motion of an electron in a Penning trap
\cite{BG}. The electron is subject to a uniform magnetic field, as before, and
in addition to the electrostatic quadrupol potential
\begin{equation}\label{12}
       e \phi (x,y,z) = \frac{1}{2} \, m \omega_z^2 (- \frac{1}{2} x^2
       - \frac{1}{2} y^2 +  z^2)\,.
\end{equation}
A non--relativistic approximation suffices and in (\ref{7}) we only
keep terms to linear order. Then the in--plane and axial motion
decouple and satisfy, with ${\mathbf r}=(x,y), {\mathbf u}= (u_1, u_2)$,
\begin{eqnarray}\label{13}
       \ddot{\mathbf r} &=& \frac{1}{2}\, \omega_z^2 {\mathbf r}
	   + \omega_c {\mathbf u}^{\perp} -
       \frac{e^2}{6 \pi c^3 m} \, \{(\omega_c^2 - \frac{1}{2}\, \omega_z^2)
       {\mathbf u} + \frac{1}{2} \omega_c \omega_z^2 {\mathbf 
       r}^{\perp} \}\,,\\
       {\ddot z} &=& - \omega_z^2  z - \frac{e^2}{6 \pi c^3 m} \, \omega_z^2
       \dot z \, .\label{14}
\end{eqnarray}

The in--plane motion can be solved easily. Without friction there
are two modes with frequencies $\omega_\pm = \frac{1}{2}\, (\omega_c \pm (
\omega_c^2 - 2
\omega_z^2)^{1/2})$. For the damping coefficients first order perturbation theory is
sufficiently accurate with the result
\begin{equation}\label{15}
        \gamma_+ = \frac{e^2}{6 \pi c^3 m}\, \frac{\omega_+^3}{\omega_+ - 
		\omega_-},
        \quad \gamma_- =
        \frac{e^2}{6 \pi c^3 m}\, \frac{\omega_-^3}{\omega_- - 
        \omega_+}\, ,
\end{equation}
in agreement with a QED resonance computation \cite{BG}. Of course
with some extra effort, one could handle also nonlinear and
relativistic effects. We emphasize that  (\ref{15}) is beyond
the capability of Larmor's formula, which works only for a
single mode.

Note that $\omega_- < \omega_+$ and therefore $\gamma_- < 0$ which
reflects that a loss of energy due to radiation lowers the
potential energy and increases the orbit size.
In practice $\omega_c/2 \pi = 164\,$GHz and  $\omega_z /2 \pi = 62
\,$MHz. Inserting in (\ref{15}) leads to $\gamma_+^{-1} = 8 \times 10^{-2}$
sec and $\gamma_-^{-1} = - 3 \times 10^{14}$ sec. Experimentally, one 
observes that the magnetron motion ($\omega_{-}$) is stable over 
weeks, whereas the cyclotron motion ($\omega_{+}$) decays to 
equilibrium within fractions of a second.

To summarize, we have investigated the solution flow
of the Lorentz--Dirac equation and discovered that  in its structure
it is identical
to renormalization group flows in critical
phenomena. The physical solutions are on the critical manifold and
are governed there by an effective second order equation. This
equation is not plagued by the difficulties usually associated
with the Lorentz--Dirac equation. In particular, the solutions to
(\ref{7}) are stable and have the correct long--time behavior. Our
examples show how radiation damping can be handled systematically
and with ease. It would be of interest to have more stringent
experimental tests. E.g. one could decrease the magnetic field in
the Penning trap, so that the two modes mix better, and try to reach
the resonance point $\omega_c^2 = 2 \omega_z^2$, where the
life--times should vanish according to (\ref{15}). A further
interesting possibility is to turn the magnetic field out of the
symmetry axis.
Then all three modes mix resulting in damping coefficients which
can be understood only on the basis of (\ref{7}).\bigskip\\
\begin{center} ***\bigskip
\end{center}

I am most grateful to Fritz
Rohrlich for instructive discussions and for the hint that the
independently derived Eq. (\ref{7}) appeared in Landau
and Lifshitz already a long time ago. I am indebted to Wolfgang
Schleich for pointing out the Penning trap as an interesting 
application. I thank M. Rauscher for details on
synchroton sources.

\end{document}